\documentclass{elsarticle}
\usepackage{graphics}
\usepackage{amssymb}
\usepackage{amsmath}
\usepackage{wasysym}
\usepackage{latexsym}
\usepackage{graphicx}
\usepackage{epsfig}
\usepackage{subfigure}
\usepackage{float}
\usepackage{color}
\usepackage{lineno}
\usepackage{verbatim}
\usepackage{enumitem}
\usepackage[T1]{fontenc}

\journal{}
\begin{document}

\begin{frontmatter}

\title{The Same Problem by Different Names: Unifying Regression Dilution and Regression to the Mean}

\author[1]{Jos\'e F.   Fontanari\corref{mycorrespondingauthor}}
\cortext[mycorrespondingauthor]{Corresponding author}
\ead{fontanari@ifsc.usp.br}

\author[2,3]{Mauro Santos}
\ead{mauro.santos@uab.es}

\address[1]{Instituto de F\'{\i}sica de S\~ao Carlos, Universidade de S\~ao Paulo, 13566-590 S\~ao Carlos, S\~ao Paulo, Brazil}

\address[2]{Departament de Gen\`etica i de Microbiologia, Grup de Gen\`omica, Bioinform\`atica i Biologia Evolutiva (GBBE), Universitat Aut\`onoma de Barcelona, Spain}

\address[3]{cE3c---Centre for Ecology, Evolution and Environmental Changes \& CHANGE - Global Change and Sustainability Institute,  Lisboa, Portugal}

\begin{abstract}
Regression to the Mean (RTM) and Regression Dilution are traditionally treated as unrelated issues in the clinical and ecological literatures. In this work, we demonstrate that within a linear errors-in-variables framework where baseline variables are subject to transient temporal or measurement noise, these two phenomena share an identical underlying mathematical signature. We unify these disparate traditions by comparing specialized clinical tools, such as the Berry shrinkage correction, with standard sign-agnostic structural estimators like Major Axis (MA) and Reduced Major Axis (RMA) regression. Using an analytical framework, we evaluate the closed-form population limits and finite-sample performance of these methods across various noise-to-signal ratios and sample sizes. Our results show that the Berry method is a specialized tool designed for clinical scenarios where a 1:1 relationship is expected. However, applying it to ecological trade-offs with negative slopes can lead to severe errors. We provide maps of optimality to identify which estimator most accurately recovers the true biological signal under different conditions. By reconciling these disparate methods, we offer a principled guide for researchers to choose the correct tool based on their data's noise profile rather than their disciplinary tradition.
\end{abstract}

\end{frontmatter}

\section{Introduction}

When measuring biological or clinical systems, researchers frequently encounter two pervasive inferential challenges that are traditionally treated as distinct phenomena in their respective literatures. In the clinical and biomedical sciences, investigators grapple with Regression to the Mean (RTM)—the statistical tendency for extreme initial observations to be followed by measurements closer to the population average, a phenomenon that routinely misleads researchers into perceiving treatment effects where only random noise exists \cite{Bland_1994}. Simultaneously, in fields such as allometry and evolutionary ecology, researchers confront Regression Dilution (or measurement error bias), where unobserved noise in an independent variable systematically attenuates the slope of a regression line toward zero \cite{Fuller_1987, Carroll_1996}. While the former is historically framed as a data-sampling artifact driven by conditioning on extreme baseline values \cite{Davis_1976, Barnett_2005} and the latter as a structural model-fitting problem \cite{Warton_2006}, they exhibit a deeply intertwined relationship. When formalized mathematically, they can be shown to share an identical statistical signature under a common, bivariate errors-in-variables framework.

To establish this connection, a crucial distinction must be made regarding the chaotic operational definitions of RTM across historical and contemporary literatures. In his seminal 1886 paper, Galton initially interpreted the ``regression toward mediocrity" of hereditary height data as a directional, biological brake designed by nature to maintain population stability \cite{Galton_1886}. It was only after realizing that the phenomenon was completely reversible---that tall sons also had fathers who reverted toward the mean---that any causal explanation had to be abandoned in favor of a symmetric statistical property of bivariate distributions. If an independent predictor is fixed by the investigator or measured perfectly without error, RTM manifests phenomenologically if one selects extreme sub-samples, yet the standard Ordinary Least Squares (OLS) slope remains an entirely unbiased estimator of the underlying structural parameter--- the true, unattenuated rate of change ($\beta$) that dictates the actual physical or biological law. Because this purely distributional conditioning leaves the structural line mathematically intact, foundational econometric texts, such as Fuller \cite{Fuller_1987}, systematically avoid the term RTM to prevent conceptual confusion. However, in modern applied research—spanning biomedical test-retest protocols \cite{Berry_1984, Chuang-Stein_1993}, ecological trade-off designs \cite{Kelly_2005, Gunderson_2023}, and longitudinal economics \cite{Friedman_1992}---the term RTM is more frequently deployed to describe a fundamentally different artifact: a directional distortion driven by transient temporal fluctuations or instrumental measurement noise in the baseline predictor ($X$) itself.

A classic historical example of confusing this transient measurement noise with a structural law is Horace Secrist's 1933 study on the triumph of mediocrity in business \cite{Secrist_1933}, where temporary commercial noise was mistaken for macroeconomic convergence---a misinterpretation famously dismantled by Harold Hotel\-ling's proof of the regression fallacy \cite{Hotelling_1933}. Under this errors-in-variables formulation, exemplified by a clinical protocol where a patient's blood pressure is measured on consecutive days, selecting extreme baseline readings on Day 1 guarantees a decay toward the mean on Day 2 purely due to the random fluctuation of temporary physiological noise. It is strictly under this conditional framework---where $X$ is a stochastic variable carrying unobserved noise---that the statistical mechanism driving the artifactual RTM slope decay becomes mathematically isomorphic to regression dilution, flattening the OLS line and causing it to fail in its structural estimation goal.

This structural failure carries profound consequences for empirical inference. When noise in the predictor variable goes unaccounted for, the resulting attenuation bias does not merely introduce uncertainty; it predictably suffocates the empirical signal, leading researchers to conclude that a relationship is weak or negligible when it is, in fact, biologically decisive \cite{Reiersol_1950}.  Accurate estimation of the true $\beta$ is critical precisely because this parameter quantifies the absolute rate of change or the physical magnitude of an evolutionary constraint. Consequently, within this errors-in-variables regime, the primary goal of the regression dilution literature shifts from simple descriptive line-fitting to the deployment of alternative statistical tools capable of penetrating measurement artifacts to isolate the underlying structural law \cite{Hutcheon_2010}.

To this end, researchers in allometry and physical anthropology have long utilized symmetric estimators, such as Major Axis (MA) and Reduced Major Axis (RMA) regression, as structural alternatives to OLS \cite{Ricker_1973, Jolicoeur_1975, Smith_2009}. While these methods were historically championed for their geometric symmetry, their primary value in structural modeling---where a causal hierarchy is often assumed a priori---lies in their capacity to account for noise in the independent variable. These estimators provide a general-purpose solution for error-in-variables by assuming specific, rigid constraints on the noise distributions across the parameter space.

In contrast, other corrections have emerged from a more targeted, directional tradition. A prominent example is the estimator proposed by Berry et al. \cite{Berry_1984}, which originated in the clinical literature to specifically address RTM artifacts in test-retest scenarios. Unlike the geometric estimators, this method is designed to protect against Type I errors by anchoring the null hypothesis to $\beta = 1$, representing the assumption of biological stability between measurements. By shrinking the observed slope toward unity, it provides a conservative test for clinical intervention effects in a clearly defined causal outcome.

Controversies arise, however, when this specialized shrinkage tool is extended to fields such as behavior and ecology to purge longitudinal data of RTM effects \cite{Kelly_2005,Gunderson_2023}. While the Berry method is exceptionally stable in clinical settings where $\beta \approx 1$, it can fail catastrophically when applied to systems where the structural relationship is expected to be zero or negative. This is particularly evident in the Trade-off Hypothesis (TOH) of thermal plasticity \cite{Angilletta_2003,Santos_2025}, where the true $\beta$ defines the steepness of a performance trade-off. In such cases, using a method anchored to a $\beta = 1$ "no-change" hypothesis does not simply add noise; it systematically misrepresents the evolutionary constraints acting on the organism and may mathematically erase the biological signal entirely \cite{Fontanari_2026}. This highlights the need for a rigorous comparison between specialized clinical tools and the more flexible, sign-agnostic triad of OLS, MA, and RMA.

The principal aim of this work is therefore to provide a unifying framework that bridges these disparate literatures---clinical and ecological.  Using the structural model of Hayes \cite{Hayes_1988}, we derive the population-level analytical slopes for OLS, MA, RMA, and the Berry estimator. This mathematical foundation allows for a rigorous comparison of the inherent biases of each method across the full range of noise-to-signal ratios for both variables. Furthermore, we extend this analysis to finite samples through numerical simulations, using the Mean Squared Error (MSE) to determine the domain of optimality for each estimator. This dual approach allows us to evaluate not only the structural accuracy of these methods at the population limit but also their practical performance in the small-sample regimes typical of empirical research. In doing so, we provide a principled guide for estimator selection that transcends disciplinary boundaries and replaces heuristic choice with analytical clarity.

The remainder of this paper is structured to maintain a rigorous and transparent narrative across these multi-disciplinary connections, organized around a clear tripartite framework. We begin by establishing the mathematical equivalence between RTM and attenuation bias; specifically, in Section~\ref{sec:2}, we formalize the foundational errors-in-variables linear structural model to demonstrate how unobserved noise in the independent variable drives a shared statistical signature. We then shift to analyzing the geometric and statistical properties of the various estimators. Section~\ref{sec:3} provides the analytical derivation of the primary triad (OLS, MA, and RMA) and identifies the conditions required to recover the structural parameters. Section~\ref{sec:4} extends this by presenting a comparative analysis of their asymptotic biases, utilizing phase diagrams in the noise-to-signal parameter plane to map their relative boundaries. In Section~\ref{sec:5}, we explicitly reconcile the clinical RTM literature with this structural framework by deriving the population-level analytical form of the Berry shrinkage estimator \cite{Berry_1984}. We evaluate the impact of sampling variation on these estimators in Section~\ref{sec:6}, mapping how finite sample sizes distort their domains of optimality. Finally, Section~\ref{sec:7} addresses the practical implications for applied inference in clinical and ecological contexts, contrasting localized clinical interventions with the global demands of modeling negative biological trade-offs before offering concluding remarks.


\section{Statement of the problem}\label{sec:2}

Let $X$ and $Y$ be random variables representing the true state of a subject at two different points in time or under different conditions. In clinical research, for example, $X$ may represent a baseline physiological measurement (such as blood pressure or glucose levels) and $Y$ a subsequent measurement following a period of observation or intervention. Unlike the symmetric scaling problems common in morphology, we focus here on scenarios with a clear directional or causal hierarchy, where $X$ is the antecedent variable. We assume these true states are related by the linear structural equation
\begin{equation}\label{Y}
Y =  \alpha + \beta X,
\end{equation}
where $\alpha$ and $\beta$ are the parameters to be estimated. We model $X$ as a random variable drawn from a normal distribution, $X \sim N(\mu, \sigma^2)$, where $\mu$ is the population mean and $\sigma^2$ represents the between-subject variance. It follows that $Y$ is also a normal random variable with mean $\alpha + \beta \mu$ and variance $\beta^2 \sigma^2$.

By adopting this asymmetric framework, we focus on the specific challenges posed by measurement error in the $X$ variable, which is the primary driver of Regression to the Mean  and slope attenuation. While symmetric estimators like MA and RMA are often justified by a lack of causal direction \cite{Smith_2009}, here we evaluate them as potential tools for correcting the systematic biases introduced by baseline measurement noise. In this context, the researcher's goal is to recover the structural parameter $\beta$---representing the true biological stability or change---despite the presence of observational noise that inevitably dilutes the OLS estimate.

The difficulty is that the true values, $X$ and $Y$, are not directly observable. Instead, we measure values $x$ and $y$, which are subject to within-subject variation. We model these observed values as
\begin{eqnarray}\label{xy}
x & = & X + \epsilon  \nonumber \\
y & = & Y + \zeta,
\end{eqnarray}
\unskip
where $\epsilon$ and $\zeta$ are independent random variables representing this within-subject variation. We assume they are normally distributed, $\epsilon \sim N(0,\delta^2)$ and $\zeta \sim N(0,\nu^2)$. This formulation is identical to the linear structural model introduced by Hayes \cite{Hayes_1988} (see also \cite{Fontanari_2026}) to investigate how measurement error drives the spurious dependence of a variable's change on its initial value.

The interpretation of $\epsilon$ and $\zeta$ as pure measurement errors is standard in the physical sciences, where $X$ and $Y$ may be governed by a strict deterministic law and any deviation is attributed to imperfect instrumentation. However, as Smith \cite{Smith_2009} extensively discusses, this terminology can be an oversimplification in biological and ecological systems. Here, these terms frequently capture what is known as equation error  or natural non-structural scatter---encompassing short-term biological fluctuations, transient physiological noise (e.g., circadian metabolic shifts or seasonal fat dynamics), and micro-environmental stochasticity.

From an interpretative standpoint, treating these stochastic terms as instrumental noise versus transient biological scatter represents two distinct conceptual views. Statistically, however, as long as this short-term biological heterogeneity is uncorrelated with the latent variables and with each other, its mathematical signature and consequences on the regression estimators are strictly identical to those of classical measurement error \cite{Smith_2009}. Consequently, the parameter $\beta$ does not imply a rigid, deterministic Platonic truth.  Instead, it represents the underlying, long-term functional scaling relation or evolutionary trade-off architecture that governs the system once transient individual noise and temporal fluctuations are averaged out. We maintain this rigorous errors-in-variables framework throughout our analyses.

The slope of the linear regression of $y$ on $x$ is the OLS estimate $\beta_{OLS}$ of the true slope $\beta$. This estimate is defined as the ratio of the covariance between $y$ and $x$ to the variance of $x$ \cite{Wasserman_2004}%
\begin{equation}\label{eq:beta_{OLS}}
\beta_{OLS} = \frac{\mathrm{cov}(x,y)}{\mathrm{var}(x)} = \frac{\beta \sigma^2}{\sigma^2 + \delta^2} = \beta R, \end{equation}
where $R = \sigma^2 / (\sigma^2 + \delta^2)$ is the repeatability (or reliability) of the measurement. Biologically, $R$ represents the reliability of a single measurement as an indicator of the individual's true state. 

Equation (\ref{eq:beta_{OLS}}) explicitly reveals the attenuation bias, where the presence of measurement error ($\delta^2 > 0$) strictly biases the OLS slope toward zero. The physical meaning of this bias is governed by $R$: when measurement error is absent ($\delta^2 = 0$), $R = 1$ and the observed variable $x$ is identical to the true state $X$. However, as measurement error $\delta^2$ grows relative to the between-subject variance $\sigma^2$, the repeatability $R$ decreases, thereby obscuring the true biological signal and diluting the estimated relationship.

Similarly, the OLS  estimate of the intercept, $\alpha_{OLS}$, is derived from the standard ordinary least squares (OLS) definition, which anchors the regression line at the centroid of the observed data, $(\mu, \alpha + \beta\mu)$
\begin{equation}\label{eq:intercept_def}
\alpha_{OLS} = E[y] - \beta_{OLS} E[x].
\end{equation}
Since the error terms $\epsilon$ and $\zeta$ have zero mean, the expected values of the observed variables are identical to the expected values of the true variables: $E[x] = \mu$ and $E[y] = \alpha + \beta\mu$. Substituting these expected values  into Equation (\ref{eq:intercept_def}) yields
\begin{equation}\label{eq:alpha_c}
\alpha_{OLS}  = \alpha + \frac{\beta\mu\delta^2}{\sigma^2 + \delta^2}.
\end{equation}
Equations (\ref{eq:beta_{OLS}}) and (\ref{eq:alpha_c}) formally demonstrate that within-subject variation in the independent variable not only attenuates the slope but systematically biases the intercept. Notably, the intercept bias relies heavily on both the population mean $\mu$ and the magnitude of the measurement error $\delta^2$.  A fundamental insight from these equations is that the OLS slope and the OLS intercept are independent of the between-subject variation ($\nu^2$) on the variable $y$.

We note that the mathematical results derived here unify two phenomena often treated as distinct across scientific literatures: Regression to the Mean and Regression Dilution bias. While Galton \cite{Galton_1886} famously characterized the tendency of extreme observations to be followed by more average ones as regression towards mediocrity,  a parallel tradition in epidemiology and ecology refers to the same mathematical bias---arising from the imperfect correlation between variables—as  regression dilution \cite{Rosner_1989, MacMahon_1990}. Historically, this effect was also described in the psychological literature as attenuation \cite{Spearman_1904}, referring to the way measurement error strictly reduces the magnitude of an observed association.

Our derivation of the OLS  slope,  Equation (\ref{eq:beta_{OLS}}), makes the identity of these concepts explicit. Whether one is observing a single variable at two time points (the RTM case) or the structural relationship between two different variables (the allometric case), the presence of within-subject variation $\delta^2$ in the $x$-axis variable performs the same operation: it contracts the true structural relationship toward the null (a slope of zero). By recognizing that RTM, regression dilution, and attenuation are merely different disciplinary names for the same effect of measurement error on the OLS slope, researchers can better appreciate that the statistical remedies developed for one field are directly applicable to the others.

\section{Classic Structural Estimators}\label{sec:3}

Building upon the OLS framework established in the previous section, we now introduce the analytical forms for the structural geometric estimators---Major Axis (MA) and Reduced Major Axis (RMA). While $\beta_{OLS}$ is mathematically derived under the classical assumption that measurement error resides only in the dependent variable ($y$), our focus in Section \ref{sec:2} was to demonstrate how violating this assumption by introducing noise in $x$ leads to systematic slope attenuation. MA and RMA provide alternative structural solutions designed to account for variance in both variables simultaneously \cite{Warton_2006}. Deriving these estimators within our current framework allows us to situate the Berry correction \cite{Berry_1984} within a broader geometric context. By first establishing the mathematical properties of the standard sign-agnostic MA and RMA estimators, we demonstrate the specialized nature of the Berry correction in Section~\label{sec:5}.

In contrast, the Method of Moments estimator \cite{Carroll_1996}---also widely known as the OLS slope corrected for attenuation \cite{McArdle_2003, Smith_2009}---successfully recovers the true structural slope $\beta$, provided that the measurement error variance $\delta^2$ is known or can be independently estimated. While this correction is frequently referred to as Blomqvist’s estimator \cite{Blomqvist_1977, Chiolero_2013} within the clinical literature on test-retest change scores, it represents a fundamental structural correction applicable to any linear model. By directly scaling the observed $\beta_{OLS}$ by the inverse of the reliability ratio ($1/R$), this approach possesses a distinct practical advantage: it remains mathematically independent of the noise in the dependent variable ($\nu^2$). Consequently, it allows for a theoretically unbiased recovery of the structural relationship without requiring the researcher to quantify the error structure of both axes simultaneously, remaining perfectly valid regardless of whether $\beta$ is positive or negative.

However, the allometric and biological regression literatures have historically favored a different family of estimators to address noisy independent variables ($\delta^2 > 0$). Rather than correcting the OLS slope directly using $\delta^2$, these methods attempt to fit the structural line by making explicit assumptions about the ratio of the error variances, $\lambda = \text{var}(\zeta)/\text{var}(\epsilon) = \nu^2 / \delta^2$. The two most well-known estimators in this family---the Major Axis (MA) and Reduced Major Axis (RMA)---are not typically used in the RTM literature, and their validity (i.e., the magnitude and direction of their biases in estimating $\beta$) has not been fully assessed using a rigorous population analysis. While McArdle \cite{McArdle_1988} offered an MSE comparison of OLS, MA, and RMA using simulations of finite artificial datasets, our linear structural model allows for an exact algebraic derivation of these estimators and their inherent biases.

To evaluate the performance of these various estimators, we use the general solution for the structural slope, which requires the researcher to specify an assumed value for the ratio of the error variances, $\lambda = \nu^2 / \delta^2$. Because the true error variances are rarely known in practice, choosing $\lambda$ effectively amounts to making an educated guess or relying on methodological conventions. Imposing a value for $\lambda$ leads to a quadratic characteristic equation (see Appendix \ref{appA}). To make the sign and branch conventions explicit---particularly for instances where the underlying biological relationship represents a negative trade-off---we select the root that preserves the sign of the observed sample covariance, yielding the general Deming estimator \cite{Deming_1943,Madansky_1959}
\begin{equation}\label{eq:beta_lambda}\beta_\lambda = \frac{\mathrm{var}(y) - \lambda \mathrm{var}(x) + \sqrt{[\mathrm{var}(y) - \lambda \mathrm{var}(x)]^2 + 4 \lambda \mathrm{cov}(x,y)^2}}{2 \mathrm{cov}(x,y)}.\end{equation}
Crucially, Equation (\ref{eq:beta_lambda}) is strictly sign-agnostic. When the observed covariance is positive ($\mathrm{cov}(x,y) > 0$), the formula evaluates to the standard positive root. Conversely, when the observed covariance is negative ($\mathrm{cov}(x,y) < 0$), the positive numerator divided by the negative denominator correctly yields a negative slope ($\beta_\lambda < 0$), ensuring mathematical continuity across the entire parameter space. At the population level, substituting our structural parameters---$\mathrm{var}(x) = \sigma^2 + \delta^2$, $\mathrm{var}(y) = \beta^2\sigma^2 + \nu^2$, and $\mathrm{cov}(x,y) = \beta\sigma^2$---into Equation (\ref{eq:beta_lambda}) allows us to determine the exact asymptotic value of $\beta_\lambda$ for any assumed $\lambda$, regardless of the sign of $\beta$.

This general expression is particularly illuminating because it encompasses several common regression techniques as special cases based on the researcher's assumptions about the unobservable errors. For instance, the standard OLS slope ($\beta_{OLS}$) is recovered as $\lambda \to \infty$, effectively assuming that the independent variable is measured perfectly without error. The Major Axis (MA) slope, which minimizes the perpendicular distances from the data points to the regression line, corresponds to the guess that the error variances are strictly equal ($\lambda = 1$). The Reduced Major Axis (RMA) slope—frequently championed in allometry when both variables are subject to error—arises when one assumes that the ratio of error variances is exactly equal to the ratio of the observed phenotypic variances, $\lambda = \mathrm{var}(y)/\mathrm{var}(x)$ \cite{Smith_2009}. Unlike other approaches that demand external calibration, the RMA is often favored precisely because it bases its estimate of $\lambda$ solely on the observed variance of the data.

Of course, to use Equation (\ref{eq:beta_lambda}) to compute the exact structural slope, one must independently know the error variances of both variables ($\delta^2$ and $\nu^2$) in order to determine the correct error-variance ratio $\lambda = \nu^2/\delta^2$. As we have previously shown \cite{Fontanari_2026}, bivariate data alone result in a plateau of non-identifiability where the structural slope cannot be resolved without such external information. This strict requirement highlights the practical utility of the alternative Method of Moments estimator discussed previously. Because this method applies a direct algebraic correction to the OLS slope, viz. $\beta_{MM}= \beta_{OLS} (1+ \delta^2/\sigma^2)$, it bypasses the $\lambda$ formulation entirely and requires an estimate of only the measurement error $\delta^2$, without needing to quantify the biological noise $\nu^2$.

Once a structural slope is estimated via Equation (\ref{eq:beta_lambda}), the corresponding estimate for the intercept, $\alpha_\lambda$, is obtained by constraining the regression line to pass through the centroid of the observed data. As established previously, the expected values of the observed variables are $E[x] = \mu$ and $E[y] = \alpha + \beta\mu$. Substituting these into the standard intercept definition yields
\begin{equation}\label{eq:alpha_lambda}
\alpha_\lambda = E[y] - \beta_\lambda E[x] = (\alpha + \beta\mu) - \beta_\lambda \mu = \alpha + \mu (\beta - \beta_\lambda).
\end{equation}
An elegant consequence of this linear relationship is that the bias in the intercept is strictly proportional to the bias in the slope. If a researcher manages to select the exact true ratio $\lambda = \nu^2/\delta^2$, or similarly applies an unbiased correction like the Method of Moments such that the slope estimate is exact ($\beta_\lambda = \beta$), the intercept estimate is automatically perfectly unbiased, yielding $\alpha_\lambda = \alpha$. Conversely, any assumption about $\lambda$ that over- or underestimates the structural slope will simultaneously corrupt the intercept, with the magnitude of that secondary bias scaled by the population mean $\mu$. Because the knowledge of the slope, combined with the requirement that the regression line passes through the centroid, uniquely fixes the intercept, we shall henceforth focus our analysis exclusively on the properties and biases of the slope estimators.

\subsection{The Major Axis (MA) Estimator}

The Major Axis (MA) regression, also known as orthogonal regression, seeks to minimize the sum of the squared perpendicular distances from the observed data points to the fitted line \cite{Sokal_1995, Warton_2006}. In the context of the general structural model evaluated via maximum likelihood (often referred to as Deming regression \cite{Deming_1943}), this geometric optimization is mathematically proved to be identical to assuming that the error variances of the two variables are strictly equal, such that $\lambda = 1$ \cite{Fuller_1987, Smith_2009}.

To evaluate the validity of this estimator, we determine its asymptotic population value by substituting $\lambda = 1$ and our structural population parameters---$\mathrm{var}(x) = \sigma^2 + \delta^2$, $\mathrm{var}(y) = \beta^2\sigma^2 + \nu^2$, and $\mathrm{cov}(x,y) = \beta\sigma^2$---into the general sign-agnostic formulation in Equation (\ref{eq:beta_lambda}). This substitution yields the population MA slope
\begin{equation}\label{eq:beta_MA}\beta_{MA} = \frac{\left[\sigma^2(\beta^2 - 1) + (\nu^2 - \delta^2)\right] + \sqrt{\left[\sigma^2(\beta^2 - 1) + (\nu^2 - \delta^2)\right]^2 + 4\beta^2\sigma^4}}{2\beta\sigma^2}.
\end{equation}
This equation is highly revealing and holds universally regardless of the sign of $\beta$. It shows that the MA estimator successfully recovers the true structural slope ($\beta_{MA} = \beta$) if and only if the assumption of equal error variances is exactly met in reality; that is, when $\nu^2 = \delta^2$. Under this specific condition, the terms involving the error variances cancel out, the expression under the square root simplifies to $\sigma^4(\beta^2 + 1)^2$, and the entire equation perfectly reduces to $\beta$. Crucially, when $\beta < 0$, the denominator naturally carries the negative sign of the covariance, ensuring that $\beta_{MA}$ correctly preserves the negative direction of the underlying biological relationship.

However, when this assumption is violated ($\nu^2 \neq \delta^2$), the MA slope becomes systematically biased. The direction and magnitude of this bias are determined by the relationship between the error variances $\nu^2$ and $\delta^2$. In typical empirical studies, there is no a priori reason to expect the noise on the dependent variable ($\nu^2$) to perfectly match the noise on the independent variable ($\delta^2$). These errors often arise from distinct processes---whether they be instrumental limitations, physiological fluctuations, or environmental stochasticity---and their relative magnitude is rarely known in the absence of external calibration.

For example, if the noise in the dependent variable is substantially larger than the noise in the independent variable ($\nu^2 > \delta^2$), Equation (\ref{eq:beta_MA}) shows that the MA estimator will systematically overestimate the magnitude of the true slope $\beta$. Conversely, if the noise in $x$ dominates ($\delta^2 > \nu^2$), the MA slope will be biased toward zero, suffering from a residual dilution effect. Because the MA estimator fundamentally relies on the assumption of equal error variances rather than an empirical calibration of those errors, its application to unstandardized data is highly susceptible to severe, unpredictable biases.

\subsection{The Reduced Major Axis (RMA) Estimator}

The Reduced Major Axis (RMA) estimator, sometimes referred to as the standard major axis or geometric mean regression, is exceptionally popular in allometry and comparative biology \cite{Kermack_1950, Sokal_1995,Warton_2006}. Geometrically, it minimizes the sum of the areas of the right triangles formed by the data points and the regression line. Within our $\lambda$ framework, this geometric optimization is mathematically equivalent to assuming that the ratio of the error variances is exactly equal to the ratio of the total observed phenotypic variances, namely $\lambda = \mathrm{var}(y)/\mathrm{var}(x)$ \cite{McArdle_1988, Smith_2009}.

The unique algebraic property of the RMA estimator becomes immediately apparent when we substitute this specific assumption into the general structural Equation (\ref{eq:beta_lambda}). Because $\lambda \mathrm{var}(x)$ perfectly cancels out $\mathrm{var}(y)$, the term $\mathrm{var}(y) - \lambda \mathrm{var}(x)$ becomes zero. To make the sign convention explicit across all domains, we simplify the square root of the squared covariance term by introducing the sign function, $\sqrt{\mathrm{cov}(x,y)^2} = \mathrm{cov}(x,y) \cdot \mathrm{sgn}[\mathrm{cov}(x,y)]$. The equation then simplifies cleanly to
\begin{equation}\label{eq:beta_rma_simplification}
\beta_{RMA} = \frac{\sqrt{4 \left(\frac{\mathrm{var}(y)}{\mathrm{var}(x)}\right) \mathrm{cov}(x,y)^2}}{2 \mathrm{cov}(x,y)} = \mathrm{sgn}[\mathrm{cov}(x,y)] \sqrt{\frac{\mathrm{var}(y)}{\mathrm{var}(x)}}.\end{equation}
To determine the validity of the RMA estimator at the population level, we substitute our true structural parameters into Equation (\ref{eq:beta_rma_simplification}). Noting that 
$$\mathrm{sgn}[\mathrm{cov}(x,y)] = \mathrm{sgn}[\beta\sigma^2] = \mathrm{sgn}(\beta),$$
and using $\mathrm{var}(x) = \sigma^2 + \delta^2$ and $\mathrm{var}(y) = \beta^2\sigma^2 + \nu^2$, we obtain the asymptotic, sign-preserving RMA slope
\begin{equation}\label{eq:beta_RMA_structural}
\beta_{RMA} = \mathrm{sgn}(\beta) \sqrt{\frac{\beta^2\sigma^2 + \nu^2}{\sigma^2 + \delta^2}} = \beta \sqrt{\frac{1 + \frac{\nu^2}{\beta^2\sigma^2}}{1 + \frac{\delta^2}{\sigma^2}}}.
\end{equation}
Equation (\ref{eq:beta_RMA_structural}) provides a strict mathematical diagnosis of the RMA estimator's inherent bias. The RMA succeeds in recovering the exact structural slope ($\beta_{RMA} = \beta$) under one highly specific condition: the term inside the square root must equal 1. This occurs only when $\nu^2 / (\beta^2\sigma^2) = \delta^2 / \sigma^2$, or equivalently, when the error variances are proportional to the structural signal on each axis, such that $\nu^2 = \beta^2 \delta^2$. In other words, the signal-to-noise ratio must be identical for both measured variables.

In biological reality, this is a remarkably stringent and generally unjustifiable assumption. In the context of allometric scaling, for example, the error variance of the metabolic rate measurement ($\nu^2$)---which is influenced by post-absorptive state, circadian rhythms, and technical gas-exchange precision---is governed by entirely different processes than the error variance of the body mass measurement ($\delta^2$), which may be subject to gut content or seasonal fluctuations \cite{White_2003}. There is no a priori reason to assume these disparate noise sources scale perfectly with the structural signal on each axis. When this proportionality fails, the RMA estimator becomes systematically biased.

Crucially, unlike the OLS slope which is strictly attenuated toward zero by the noise in $x$ ($\delta^2$), Equation (\ref{eq:beta_RMA_structural}) reveals that the RMA slope can be biased in either direction. If the noise in the dependent variable is proportionally larger than the noise in the independent variable relative to the squared slope ($\nu^2 > \beta^2\delta^2$), the RMA will systematically overestimate the true structural relationship. By forcing the estimator to absorb the total variance of $y$ into the slope calculation, the RMA fundamentally confounds the noise in $y$ ($\nu^2$) with the structural relationship ($\beta$), thereby inflating the estimated slope.

\subsection{The Exact Solution: Recovering the True Structural Slope}

After identifying the inherent biases of the MA and RMA estimators, we must confirm that the $\lambda$ framework is internally consistent. Specifically, we show that if a researcher possesses independent knowledge of the error variances ($\delta^2$ and $\nu^2$)---thereby setting $\lambda$ exactly equal to the true ratio $\nu^2/\delta^2$---Equation (\ref{eq:beta_lambda}) flawlessly recovers the true structural slope $\beta$.

By substituting the true parameters $\mathrm{var}(x) = \sigma^2 + \delta^2$, $\mathrm{var}(y) = \beta^2\sigma^2 + \lambda\delta^2$, and $\mathrm{cov}(x,y) = \beta\sigma^2$ into the general estimator $\beta_\lambda$, the term within the square root simplifies through the completion of a perfect square
\begin{equation}\label{eq:square_root_short}(\mathrm{var}(y) - \lambda \mathrm{var}(x))^2 + 4 \lambda \mathrm{cov}(x,y)^2 = \left[\sigma^2(\beta^2 - \lambda)\right]^2 + 4\lambda\beta^2\sigma^4 = \sigma^4(\beta^2 + \lambda)^2.
\end{equation}
Substituting this result back into the full estimator yields
\begin{equation}\label{eq:exact_beta_short}\beta_\lambda = \frac{\sigma^2(\beta^2 - \lambda) + \sigma^2(\beta^2 + \lambda)}{2\beta\sigma^2} = \beta.
\end{equation}
Because the numerator simplifies strictly to $2\beta^2\sigma^2$, dividing by the denominator $2\beta\sigma^2$ returns the true structural slope $\beta$. This algebraic identity holds perfectly for both positive and negative values of $\beta$, confirming that the general Deming framework \cite{Deming_1943} successfully uncovers the underlying structural signal across the entire parameter space when the true error ratio $\lambda$ is known.

However, this mathematical elegance underscores the fundamental epistemological limit of bivariate analysis: the true ratio $\lambda$ cannot be derived from sample variances alone. As highlighted by the non-identifiability plateau \cite{Fontanari_2026}, achieving this exact solution requires independent estimates of the noise in both variables. It is worth noting, however, that if the researcher can quantify the noise in $x$ ($\delta^2$) alone, the Method of Moments estimator \cite{Carroll_1996} can recover $\beta$ without requiring any knowledge of $\nu^2$. In practice, quantifying the measurement repeatability of the independent variable remains the single most effective intervention for eliminating the Regression to the Mean artifact \cite{Fontanari_2026}.

\section{Comparative Analysis of the Estimators}\label{sec:4}

Having established the exact population limits for the OLS ($\beta_{OLS}$), Major Axis ($\beta_{MA}$), and Reduced Major Axis ($\beta_{RMA}$) estimators, we now turn to a comparative analysis of their relative biases. To facilitate an analytical comparison, it is mathematically convenient to define dimensionless noise-to-signal ratios for both variables. Let $\tau_x = \delta^2/\sigma^2$ represent the relative noise in the independent variable, and let $\tau_y = \nu^2/\sigma^2$ represent the relative noise in the dependent variable.

Using these dimensionless parameters, the three asymptotic estimators can be elegantly rewritten as functions of the true slope $\beta$ and the noise-to-signal ratios:
\begin{eqnarray}
\beta_{OLS} & = & \beta \left( \frac{1}{1 + \tau_x} \right) \label{eq:comp_ols} \\
\beta_{RMA} & = & \beta \sqrt{\frac{1 + \tau_y/\beta^2}{1 + \tau_x}} \label{eq:comp_rma} \\
\beta_{MA} & = & \frac{(\beta^2 - 1) + (\tau_y - \tau_x) + \sqrt{[(\beta^2 - 1) + (\tau_y - \tau_x)]^2 + 4\beta^2}}{2\beta} .\label{eq:comp_ma}
\end{eqnarray}
Expressing the estimators in this format immediately reveals several fundamental analytical properties regarding their biases that hold universally, regardless of the sign of the structural relationship. First, Equation (\ref{eq:comp_ols}) confirms that the OLS estimator strictly attenuates the true slope toward zero ($|\beta_{OLS}| < |\beta|$) for any non-zero noise in the independent variable ($\tau_x > 0$), completely independent of the noise in $y$ ($\tau_y$). This attenuation manifests as an underestimation of the magnitude of the slope, whether the underlying biological relationship is positive or represents a negative trade-off.

Second, because the noise in the dependent variable adds strictly positive variance ($\tau_y > 0$), a direct comparison of Equations (\ref{eq:comp_ols}) and (\ref{eq:comp_rma}) proves that the magnitude of the RMA slope is strictly greater than that of the OLS slope ($|\beta_{RMA}| > |\beta_{OLS}|$). The RMA estimator successfully corrects for attenuation bias if and only if the relative noises are balanced such that $\tau_y/\beta^2 = \tau_x$. However, if the noise in $y$ is large relative to the noise in $x$ ($\tau_y/\beta^2 > \tau_x$), the RMA will systematically overcorrect, resulting in an inflated slope magnitude ($|\beta_{RMA}| > |\beta|$).

Similarly, the MA estimator explicitly depends on the absolute difference between the noise-to-signal ratios, $\tau_y - \tau_x$. If $\tau_y = \tau_x$, the term vanishes and the MA estimator perfectly recovers $\beta$. However, when the vertical noise dominates ($\tau_y > \tau_x$), the MA estimator will overcorrect and overestimate the magnitude of the structural slope. When $\beta < 0$, the denominator in Equation (\ref{eq:comp_ma}) naturally carries the negative sign, ensuring that both overestimation and underestimation dynamics preserve the correct directional branch of the relationship.

An important analytical question is whether the corrected structural estimators (MA and RMA) always outperform the theoretically flawed OLS slope. While $\beta_{OLS}$ is strictly attenuated by regression dilution, both the MA and RMA can suffer from regression inflation---a systematic overestimation of the slope's magnitude---if $\tau_y$ is sufficiently large. Specifically, the RMA estimator becomes more biased than the uncorrected OLS (i.e., $|\beta_{RMA} - \beta| > |\beta - \beta_{OLS}|$) when the noise in $y$ heavily dominates the noise in $x$. In such regimes, the attempt to "correct" for RTM via RMA is counterproductive, as the resulting overcorrection introduces a larger absolute error than the original dilution bias.

To systematically map the regimes where each estimator is subject to these competing biases, we perform a comparative evaluation of the asymptotic expressions derived in Equations (\ref{eq:comp_ols})–(\ref{eq:comp_ma}). For a fixed true structural slope $\beta$, we identify the dominance regions in the $(\tau_x, \tau_y)$ plane. Crucially, because the true position of an empirical dataset within this plane cannot be identified from bivariate data alone \cite{Fontanari_2026}, these regions should not be viewed as a mechanical decision-matrix for estimator selection. Instead, they serve as asymptotic maps of structural sensitivity, illustrating how vulnerable each geometric technique is to violations of its intrinsic, unverified assumptions about the error structure.

\subsection{Pathological Behavior under the Null Hypothesis}

Before proceeding to a numerical evaluation, it is instructive to consider the pathological case where there is no true biological relationship between the variables, such that the structural slope is exactly zero ($\beta = 0$). Evaluating the estimators under this null condition exposes a severe vulnerability in the structural correction methods that rely on the $\lambda$ framework.

From Equation (\ref{eq:comp_ols}), it is clear that if $\beta = 0$, the OLS slope strictly evaluates to $\beta_{OLS} = 0$. Because OLS relies directly on the covariance, which is exactly zero in the theoretical absence of a true relationship, OLS correctly identifies the null hypothesis regardless of the magnitude of $\tau_x$.

Conversely, the RMA estimator fails catastrophically under the null condition. Taking the limit as $\beta \to 0$, the theoretical magnitude of the RMA slope becomes entirely dictated by the ratio of the independent noises
\begin{equation}
|\beta_{RMA}| = \sqrt{\frac{\tau_y}{1 + \tau_x}}.
\end{equation}
Because the RMA forces a structural relationship based on the total variances rather than the covariance, any residual noise in the dependent variable ($\nu^2 > 0$) will cause the RMA to estimate a non-zero slope even in the absence of a true relationship.  In empirical applications, where finite sampling ensures the sample covariance is rarely perfectly zero, the RMA relies on the sign of this negligible covariance to dictate the direction of the line. Consequently, the RMA will reliably generate false positive scaling relationships out of pure noise clouds.

Similarly, the MA estimator becomes geometrically unstable. When $\beta = 0$, the true bivariate data cloud forms an orthogonal ellipse centered on the axes. If the horizontal variance exceeds the vertical variance ($1 + \tau_x > \tau_y$), the MA correctly identifies a slope of zero, as the major axis of the ellipse is flat. However, if the vertical biological scatter happens to be larger than the horizontal variance ($\tau_y > 1 + \tau_x$), the major axis of the ellipse strictly aligns with the $y$-axis. In this scenario, the MA estimator will abruptly predict a vertical line with an infinite slope. This pathological behavior strongly contraindicates the use of both MA and RMA for exploratory analyses where the existence of a true underlying structural relationship has not been firmly established \textit{a priori}.

\subsection{Domains of Optimality: Phase Diagram Analysis}\label{sec:Domains3}

\begin{figure}[t]
\center
 \includegraphics[width=1\columnwidth]{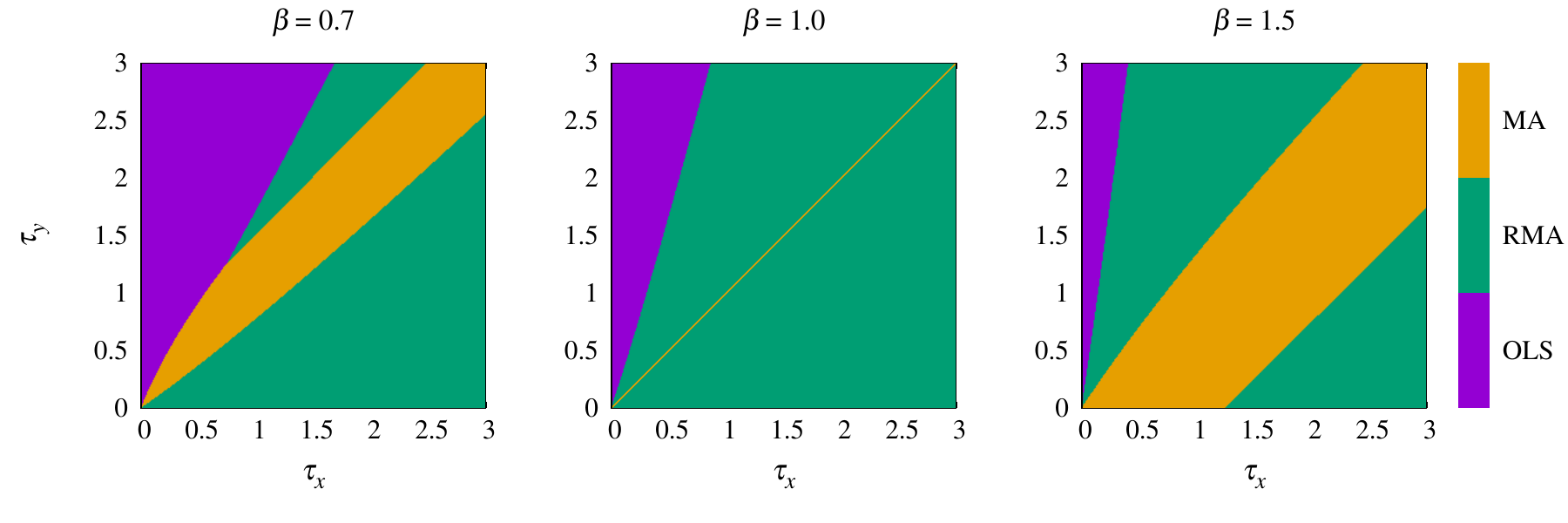}
\caption{Domains of optimality for the OLS, MA, and RMA estimators in the $(\tau_x, \tau_y)$ parameter plane, where $\tau_x = \delta^2 / \sigma^2$ and $\tau_y = \nu^2 / \sigma^2$ represent the noise-to-signal ratios of the independent and dependent variables, respectively. Results are shown for three structural slopes: $\beta=0.7$ (left), $\beta=1.0$ (center), and $\beta=1.5$ (right). The boundaries separating the OLS region from the RMA and MA regions are defined analytically by Eqs.~(\ref{eq:frontier_ols_rma}) and (\ref{eq:frontier_ols_ma}). Note the appearance of a triple point for $\beta=0.7$ and the collapse of the MA domain to the diagonal $\tau_y = \tau_x$ when $\beta=1.0$.  }
\label{fig:1}
\end{figure}

To systematically compare the estimators, Figure~\ref{fig:1} maps their domains of optimality across the $(\tau_x, \tau_y)$ parameter space for fixed structural slopes of $\beta=0.7$, $1.0$, and $1.5$. For every coordinate pair, the phase diagram identifies the estimator—OLS, MA, or RMA—that minimizes the absolute bias $|\beta_{est} - \beta|$. It is important to note that these phase boundaries are strictly invariant to the sign of the true slope. Because Equations (\ref{eq:comp_ols}), (\ref{eq:comp_rma}), and (\ref{eq:comp_ma}) dictate that a transformation from $\beta$ to $-\beta$ simply changes the sign of the resulting estimate, the absolute error $|-\beta_{est} - (-\beta)|$ reduces to $|\beta_{est} - \beta|$. Thus, the geometric regions where each estimator outperforms the others are identical for both positive and negative relationships.

As suggested by our analysis of the null case ($\beta \to 0$), the OLS estimator dominates the parameter space in the low-slope regime, particularly where $\tau_y > \tau_x + 1$. In this regime, the OLS remains stable near zero, while the MA estimator undergoes a dramatic bifurcation: it correctly recovers the null slope only when noise in the independent variable dominates ($\tau_x + 1 > \tau_y$), but diverges toward infinity when noise in the dependent variable is excessive. Consequently, the OLS serves as the most reliable default for weak structural relationships. By construction, the MA estimator is superior primarily in the vicinity of the diagonal $\tau_y = \tau_x$---a universal property arising from its assumption of equal error variances. A notable singularity occurs at $\beta=1$, where the RMA and MA estimators yield identical estimates along this diagonal, effectively collapsing the MA's independent domain of optimality. To move beyond these qualitative observations, it is instructive to derive the analytical expressions for the frontiers separating these regions.

\subsubsection{The OLS-RMA Frontier}

The boundary separating the regions where OLS and RMA are respectively optimal is defined by the condition where their absolute biases are equal. Given that $\beta_{OLS}$ strictly underestimates the true slope ($\beta_{OLS} < \beta$) and $\beta_{RMA}$ typically overestimates it in this region ($\beta_{RMA} > \beta$), the frontier is determined by the tie-breaking condition
\begin{equation}
\beta - \beta_{OLS} = \beta_{RMA} - \beta \implies \beta_{OLS} + \beta_{RMA} = 2\beta.
\end{equation}
By substituting the expressions from Eqs.~(\ref{eq:comp_ols}) and (\ref{eq:comp_rma}) into this condition, we obtain
\begin{equation}
\beta \left( \frac{1}{1 + \tau_x} \right) + \beta \sqrt{\frac{1 + \tau_y/\beta^2}{1 + \tau_x}} = 2\beta.
\end{equation}
Dividing by $\beta$ and isolating the radical yields
\begin{equation}
\sqrt{\frac{1 + \tau_y/\beta^2}{1 + \tau_x}} = 2 - \frac{1}{1 + \tau_x} = \frac{2\tau_x + 1}{1 + \tau_x}.
\end{equation}
Squaring both sides and solving for $\tau_y$ leads to the explicit analytical expression for the OLS-RMA frontier
\begin{equation}\label{eq:frontier_ols_rma}
\tau_y = \beta^2 \tau_x \left( \frac{4\tau_x + 3}{\tau_x + 1} \right). 
\end{equation}
This result shows that for very small measurement error in $x$ ($\tau_x \ll 1$), the frontier behaves as $\tau_y \approx 3\beta^2 \tau_x$, whereas for large $\tau_x$, the frontier approaches a linear slope of $4\beta^2$. This curve defines the upper boundary of the OLS dominance region in Figure~\ref{fig:1}:  above this line, the conservative nature of the OLS bias is preferable to the amplified volatility of the RMA estimate.

\subsubsection{The OLS-MA Frontier}

The Major Axis estimator $\beta_{MA}$ is the specific case of the general structural estimator $\beta_\lambda$ where $\lambda=1$. As shown in Appendix \ref{appA}, it must satisfy the characteristic quadratic equation
\begin{equation}\label{car_eq}
\beta \beta_{MA}^2 - [(\beta^2 - 1) + (\tau_y - \tau_x)]\beta_{MA} - \beta = 0.
\end{equation}
The estimator $\beta_{MA}$ corresponds to the positive root of this expression.
The frontier where OLS and MA are equally biased occurs when $\beta - \beta_{OLS} = \beta_{MA} - \beta$, which implies $\beta_{MA} = \beta (2\tau_x + 1)/(\tau_x + 1)$. Substituting this expression into the quadratic equation and solving for $\tau_y$ yields the analytical boundary
\begin{equation}\label{eq:frontier_ols_ma}
\tau_y = \tau_x \left[ 1 + \frac{\beta^2}{\tau_x + 1} + \frac{1}{2\tau_x + 1} \right]. 
\end{equation}
Taken together with Equation (\ref{eq:frontier_ols_rma}), we observe that both the OLS–RMA and OLS–MA frontiers define the boundary of OLS dominance. Because the OLS estimator effectively ignores noise in the dependent variable, its bias remains invariant to changes in $\tau_y$. In contrast, the MA and RMA estimators are highly sensitive to $\tau_y$, becoming increasingly biased as the noise in $y$ grows. Consequently, the region of OLS optimality always lies above these frontiers in the $(\tau_x, \tau_y)$ plane---representing the regimes where the noise in the dependent variable is sufficiently large to make  corrected structural estimators more biased than the original uncorrected slope.

\subsubsection{The Triple Point}

A remarkable feature of the phase diagram is the existence of a triple point where the OLS, RMA, and MA regions meet at a single coordinate $(\tau_x^*, \tau_y^*)$. At this point, all three estimators yield the same absolute error. By equating the two frontiers, Equations (\ref{eq:frontier_ols_rma}) and (\ref{eq:frontier_ols_ma}), we find that the intersection occurs when
\begin{equation}
\beta(2\tau_x + 1) = \tau_x + 1 \implies \tau_x^* = \frac{1-\beta}{2\beta - 1}.
\end{equation}
This derivation reveals a fundamental constraint: a physical triple point ($\tau_x^* > 0$) exists if and only if $0.5 < |\beta| < 1$.

For $\beta = 0.7$ (left panel of Figure~\ref{fig:1}), the triple point is located at $(\tau_x, \tau_y) = (0.75, 1.26)$. The existence of this point marks a qualitative reordering of estimator performance: for $\tau_x < \tau_x^*$, an increase in vertical noise $\tau_y$ leads to the transition sequence $\text{RMA} \to \text{MA} \to \text{OLS}$. However, for $\tau_x > \tau_x^*$, the diagram exhibits a more complex behavior where the sequence shifts to $\text{RMA} \to \text{MA} \to \text{RMA} \to \text{OLS}$, resulting in a  re-entrant $ \text{RMA} $ region. 

The geometry of these regions is highly sensitive to the structural slope. When $|\beta| > 1$, the triple point disappears from the positive quadrant, and the sequence of optimality becomes invariant across the $\tau_x$ domain (as seen in the right panel of Figure~\ref{fig:1}). A special singularity occurs at $\beta=1$ (middle panel of Figure~\ref{fig:1}).  Here, the RMA and MA estimators yield identical results along the diagonal $\tau_y = \tau_x$, causing the MA dominance region to collapse. In this case, the diagram is effectively partitioned only between the OLS and RMA regions, with the MA estimator being optimal only in the limit of the diagonal.

Conversely, as $\beta$ decreases toward the null case, the triple point is pushed toward larger values of $\tau_x$. In the limit $|\beta| \leq 0.5$, the triple point moves to infinity ($\tau_x^* \to \infty$). In this low-slope regime, the frontier between OLS and RMA vanishes entirely, and the MA region acts as a permanent buffer separating the OLS and RMA domains for all finite noise levels.

\subsubsection{The RMA-MA Frontier}

The boundary between the RMA and MA regions is mathematically more intricate than the OLS frontiers. It is defined by the condition where their absolute biases are equal,  $|\beta - \beta_{RMA}| = |\beta - \beta_{MA}|$. For $\beta = 1$, this condition simplifies to the diagonal $\tau_y = \tau_x$, where both estimators are identical and unbiased.

However, for $\beta \neq 1$, the non-linear dependence of the MA estimator on the noise ratio leads to a non-monotonic relationship between the estimators' performance. Specifically, for $\beta = 0.7$ and $\tau_x > \tau_x^*$, the equation $|\beta - \beta_{RMA}| = |\beta - \beta_{MA}|$ yields two distinct solutions for $\tau_y$. This creates the  wedge structure seen in Figure~\ref{fig:1}, where  the RMA estimator is optimal for very low $\tau_y$, yields to the MA estimator as vertical noise increases, but then becomes optimal again for a higher range of $\tau_y$ before both are eventually surpassed by the OLS. This re-entrant behavior confirms that when the assumption $\lambda=1$ is violated, the Major Axis estimator is only reliable within a specific window of noise ratios, even when compared to another bi-directional estimator like the RMA.

\section{Reconciling Clinical RTM Corrections with Structural Estimators}\label{sec:5}

While the previous sections analyzed the geometric properties of classical structural estimators, we now apply this framework to reconcile the specialized corrections developed within the clinical RTM tradition. Specifically, we focus on the Berry estimator \cite{Berry_1984}, which remains the standard tool for assessing differential drug effects in the presence of baseline-dependent noise \cite{Chuang-Stein_1993}.  This estimator was derived specifically to address clinical  test-retest artifacts and its performance has never been formally mapped against the structural benchmarks of  OLS, MA and RMA (see, however, \cite{Fontanari_2026} for a comparison with OLS in the context of change-score analysis).  By deriving the population-level form of the Berry estimator, we can now locate it within the $(\tau_x, \tau_y)$ plane and determine the specific error structures under which it provides a valid structural correction.

In the standard clinical setup, a baseline measure $x$ is taken before treatment, followed by a post-treatment measure $y$. The null hypothesis of  no differential effect implies that the treatment shifts the population mean but does not alter the underlying structural relationship between subjects. Mathematically, this corresponds to the assumption that the true structural slope is unity ($\beta = 1$). If the treatment reduces everyone’s blood pressure by exactly 10 units, the structural relationship remains a 1-to-1 correspondence.

However, as we have demonstrated in the preceding sections, the presence of measurement error or biological fluctuation in the baseline measurement $x$ induces an attenuation of the OLS slope toward zero. In a clinical context, an OLS slope $\beta_{OLS} < 1$ is frequently misinterpreted as a biological finding: it appears to show that patients with high baseline values dropped more significantly than those with low baseline values. Without accounting for the structural noise, researchers are at constant risk of mistaking a statistical artifact---Regression to the Mean---for a targeted therapeutic breakthrough. It was within this specific high-stakes environment that Berry et al.  \cite{Berry_1984} proposed a specialized correction to ``un-bias" the OLS result, unknowingly creating a bridge to the geometric estimators discussed in this work.

To address the methodology proposed by Berry et al. \cite{Berry_1984}, we must first examine the specific adjusted variable they introduced to decouple the true structural change from the noise-induced artifact. While their original derivation focused on the change score $(y - x)$, their prescription for correcting the post-treatment value $y$ can be expressed as a linear adjustment based on the reliability of the baseline measure.

By using the correlation 
\begin{equation}
\rho = \frac{\text{cov}(x,y)}{\sqrt{\text{var}(x)\text{var}(y)}} =
  \frac{\beta}{\sqrt{(1 + \tau_x)(\beta^2 + \tau_y)}}
\end{equation}
as a proxy for  the  repeatability $R=  1/(1+\tau_x)$, Berry et al. proposed that the observed post-treatment value should be adjusted relative to the group mean $E(x)$. Using the notation established in the preceding sections, the Berry-adjusted post-treatment variable $y_B$ is given by
\begin{equation}
y_B = y + (1 - \rho) [x - E(x)].
\end{equation}
This formulation is revealing. The term $(1 - \rho)$ represents the  unreliable portion of the baseline measurement---the noise that causes the observed value to deviate from the true latent mean. By adding this fraction of the baseline deviation back to the post-treatment measure $y$, the estimator effectively pushes back against the downward pressure of the RTM effect.

The resulting Berry estimator, $\beta_B$, is defined as the slope of the linear regression of $y_B$ on the original baseline $x$. Because the constant mean $E(x)$ does not affect the covariance, the derivation of this slope follows directly from the properties of the covariance operator
\begin{equation}
\beta_B = \frac{\text{cov}(y + (1 - \rho)x, x)}{\text{var}(x)} = \frac{\text{cov}(y, x)}{\text{var}(x)} + (1 - \rho).
\end{equation}
This leads to a strikingly simple relationship between the Berry estimator and the standard Ordinary Least Squares result
\begin{equation}
\beta_B = \beta_{OLS} + (1 - \rho).
\end{equation}
In this light, the Berry correction can be viewed as an additive boost to the OLS slope. Since $\rho$ is strictly less than or equal to 1, the term $(1 - \rho)$ is always non-negative, ensuring that $\beta_B$ is always greater than or equal to $\beta_{OLS}$. For the clinician, this additive constant serves as a mathematical shield,  since it compensates for the attenuation bias just enough to ensure that if the true relationship is 1:1, the resulting estimate $\beta_B$ will correctly reflect that unity, regardless of the amount of noise present in the data.

The true nature of the Berry estimator is revealed when we translate it into the geometric framework established earlier in this work. By substituting the standard identity $\beta_{OLS} = \rho \beta_{RMA}$ into the Berry slope formula, we obtain
\begin{equation}
\beta_B = \rho \beta_{RMA} + 1 - \rho =  1 + \rho (\beta_{RMA} - 1).
\end{equation}
This identity demonstrates that the Berry  estimator is fundamentally a shrinkage estimator anchored to unity. It functions by taking the Reduced Major Axis (RMA) estimate---the slope that accounts for both horizontal and vertical variance---and linearly shrinking it toward the clinical null hypothesis of $\beta = 1$. The strength  of this pull is governed by the correlation coefficient $\rho$. When the data is perfectly clean ($\rho = 1$), the estimator returns the raw RMA slope. When the data is entirely noise ($\rho = 0$), the estimator safely collapses to its anchor point of 1.

This reveals why the Berry estimator has been so successful in clinical research \cite{Chuang-Stein_1993}. It is designed to be conservative with respect to the null hypothesis. By shrinking the estimate toward 1, it ensures that as noise increases, the probability of a Type I error (falsely claiming a differential drug effect) decreases.  The use of Berry  estimator outside the clinical context as for instance in ecology \cite{Kelly_2005,Gunderson_2023} as a  general estimator for the true slope is less well justified.

\begin{figure}[t]
\center
 \includegraphics[width=1\columnwidth]{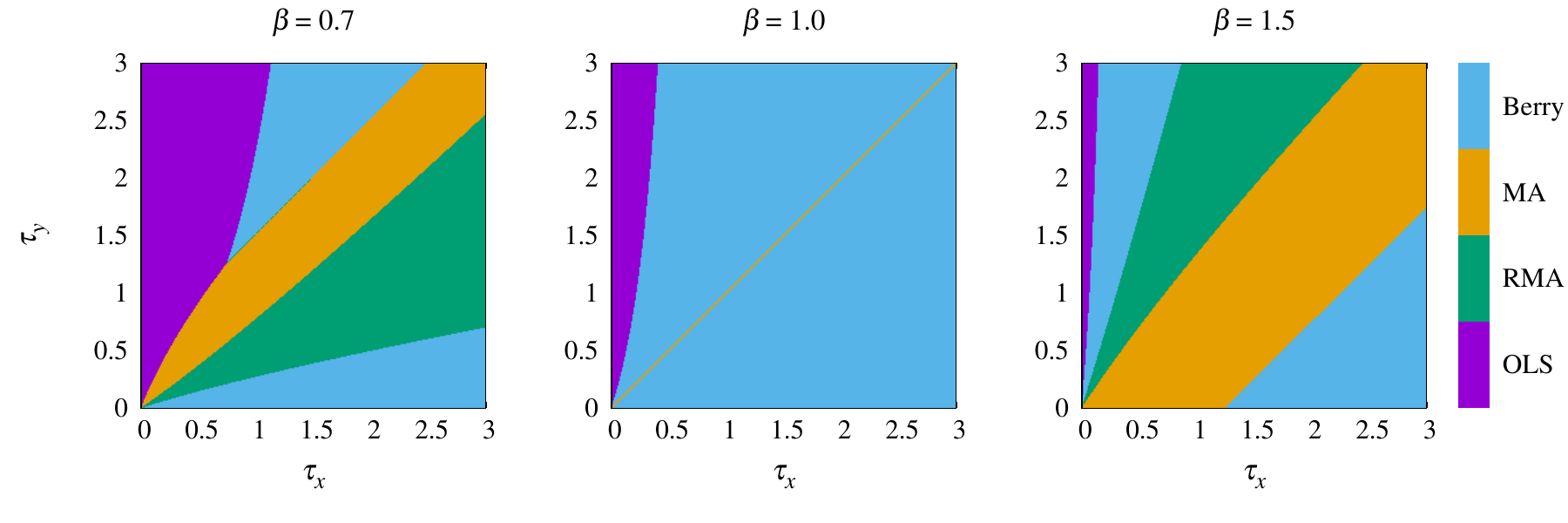}
\caption{Domains of optimality for the OLS, MA, RMA, and Berry estimators in the $(\tau_x, \tau_y)$ parameter plane. The axes represent the noise-to-signal ratios of the independent ($\tau_x = \delta^2 / \sigma^2$) and dependent ($\tau_y = \nu^2 / \sigma^2$) variables. Results are presented for three structural slopes: $\beta=0.7$ (left), $\beta=1.0$ (center), and $\beta=1.5$ (right). The dominance of the Berry estimator is concentrated around $\beta \approx 1$, reflecting its mathematical construction which intentionally biases the estimate toward unity. Note that for $\beta=1.0$, the Berry estimator dominates a large portion of the plane, as its shrinkage target coincides with the true structural value. }
\label{fig:2}
\end{figure}   
 
However, this clinical strength is also its primary limitation as a general structural tool. Unlike the OLS, MA, and RMA estimators, which are symmetric and agnostic to the sign or magnitude of the slope, the Berry estimator is asymmetric and biased toward positive relationships. In a general scientific context where the true structural relationship might be negative (e.g., $\beta = -0.7$), the Berry estimator will attempt to shrink the estimate across the origin toward $+1$, resulting in catastrophic bias.

This mathematical anchor to $+1$ explains the regions of dominance observed in the phase diagrams in Figure~\ref{fig:2}. The Berry estimator performs exceptionally well for positive structural slopes close to unity, where its inherent shrinkage toward 1 acts as a variance-reduction mechanism. However, it fails as a general-purpose tool for the broader physical and biological sciences where the null hypothesis of unity does not apply. In particular, for negative values of $\beta$, the Berry estimate is entirely supplanted by the OLS, MA, and RMA estimators.  In these cases, the resulting phase diagrams remain identical to those derived for the three-estimator comparison shown in Figure~\ref{fig:1}. Notably, the frontiers between the Berry-dominated regions and the OLS or RMA domains can be analytically evaluated  following the same procedure developed for the fundamental triad in Section \ref{sec:Domains3}.

\section{Sample Size Effects on the Domains of Optimality}\label{sec:6}

Equations (\ref{eq:comp_ols}), (\ref{eq:comp_rma}), and (\ref{eq:comp_ma}) provide the population values for the OLS, RMA, and MA regression slopes. While their simplicity allows for a complete assessment of the biases as a function of the model's parameters, a practical study relies on a finite sample of individuals. Consequently, the observed regression slopes calculated from a sample will inevitably differ from these population values due to sampling variation. In this section, we investigate the impact of this sampling variation using the Mean Squared Error (MSE). Notably, we restrict this finite-sample analysis to the OLS, MA, and RMA estimators. As argued in the previous section, the Berry estimator is a specialized shrinkage tool anchored to a specific clinical null hypothesis ($\beta=1$). Because its performance is fundamentally asymmetric and highly dependent on the proximity of the true slope to unity, it does not share the general-purpose geometric properties of the primary triad. We therefore focus our MSE analysis on the OLS, MA, and RMA estimators to provide a generalized framework for the structural modeling of trade-offs and scaling laws across the physical and biological sciences.

The MSE is a comprehensive measure of the total error of an estimator. It is mathematically defined as the expected value of the squared difference between the estimated slope ($\beta_{est}$) and the true structural parameter ($\beta$)
\begin{equation}
MSE = E[(\beta_{est}- \beta)^2].
\end{equation}
A key property of the MSE is that it can be decomposed into the sum of the estimator's squared bias and its variance
\begin{equation}
MSE = \mbox{Bias}^2 + \mbox{Variance} = (E[\beta_{est}] - \beta)^2 + \text{var}(\beta_{est}).\end{equation}
This decomposition illustrates the fundamental trade-off in statistical inference: attempts to eliminate bias (reducing the first term) often result in a significant increase in sampling variance (increasing the second term). For an applied researcher, the optimal method is typically the one that minimizes the total MSE, representing the highest probability of obtaining an estimate close to the true parameter from a single dataset. In the following analysis, we investigate how the asymptotic phase diagrams presented in Figure~\ref{fig:1} are reconfigured under finite sample size effects.

To map these finite-sample domains of optimality, we implemented a structured Monte Carlo simulation protocol across a dense grid of the noise parameter space. For each specific true structural slope ($\beta \in \{0.7, 1.5\}$), we constructed a $60 \times 60$ parameter grid where the noise-to-signal ratios $\tau_x$ and $\tau_y$ vary systematically from $0.05$ to $3.00$ in increments of $0.05$. At each specific coordinate $(\tau_x, \tau_y)$ in the parameter plane, we generated finite datasets according to the following generative mechanism:

First, a latent, unobservable structural vector of size $n$ ($n \in \{25, 50, 100\}$) was drawn from a standard normal distribution, $X \sim \mathcal{N}(0, 1)$, establishing a baseline signal variance of $\sigma^2 = 1$. Second, independent, zero-mean Gaussian measurement errors were generated using the classical Box-Muller transformation such that $\zeta \sim \mathcal{N}(0, \delta^2)$ and $\epsilon \sim \mathcal{N}(0, \nu^2)$, where the error variances were calibrated directly to the grid coordinates via $\delta = \sqrt{\tau_x}$ and $\nu = \sqrt{\tau_y}$. The observed bivariate data pairs $(x_i, y_i)$ were then assembled via $x_i = X_i + \zeta_i$ and $y_i = \beta X_i + \epsilon_i$.

For each individual grid coordinate, this stochastic data-generating process was repeated across $5 \times 10^4$ independent Monte Carlo replicates. For every replicate dataset, the sample variances and covariances were calculated using standard unbiased definitions, and the OLS, MA, and RMA slopes were computed. The empirical MSE for each estimator was tracked cumulatively across all replicates. Finally, the boundaries of the phase diagrams were mapped by identifying the specific estimator that minimized the total empirical MSE at each coordinate. Pseudorandom number generation was executed utilizing intrinsic hardware-seeded processors via standard lagged Fibonacci or congruent algorithms mapped into the Box-Muller algorithm, guaranteeing that the simulation boundaries are highly stable and fully reproducible.

\begin{figure}[ht]
\center
 \includegraphics[width=1\columnwidth]{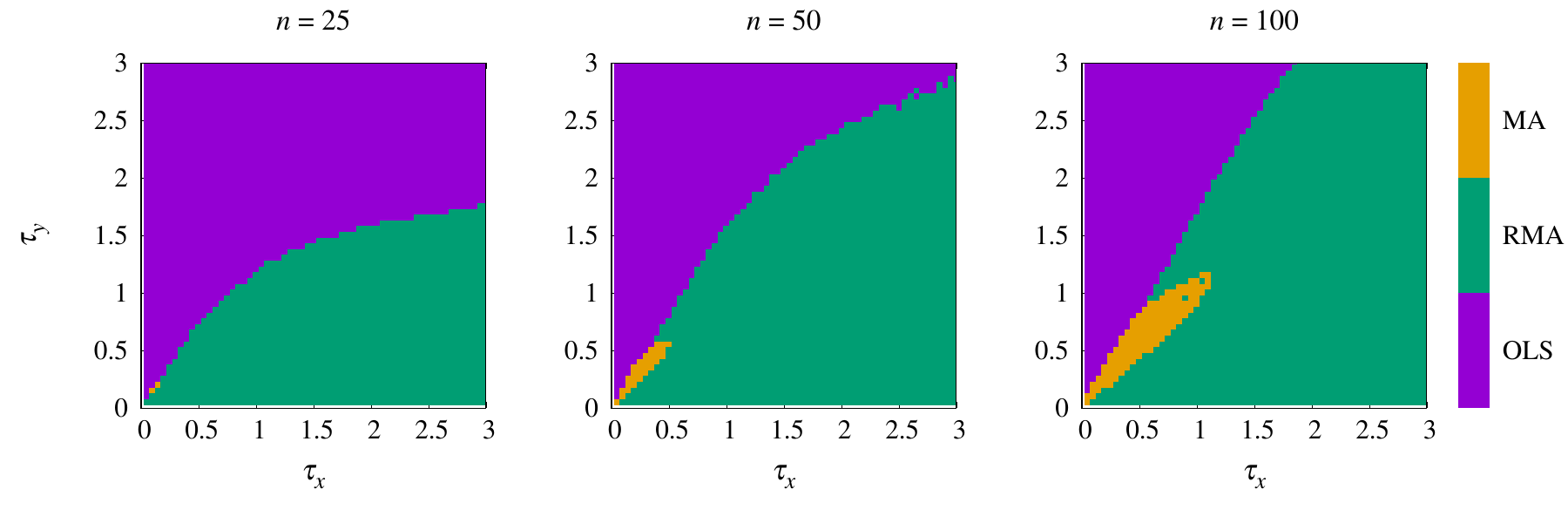}
\caption{Domains of optimality (minimum MSE) for the OLS, MA, and RMA estimators in the $(\tau_x, \tau_y)$ parameter plane for a structural slope $\beta=0.7$. The axes $\tau_x$ and $\tau_y$ represent the noise-to-signal ratios for the independent and dependent variables, respectively. Panels show results for sample sizes $n=25$ (left), $n=50$ (center), and $n=100$ (right). }
\label{fig:3}
\end{figure}   

\begin{figure}[ht]
\center
 \includegraphics[width=1\columnwidth]{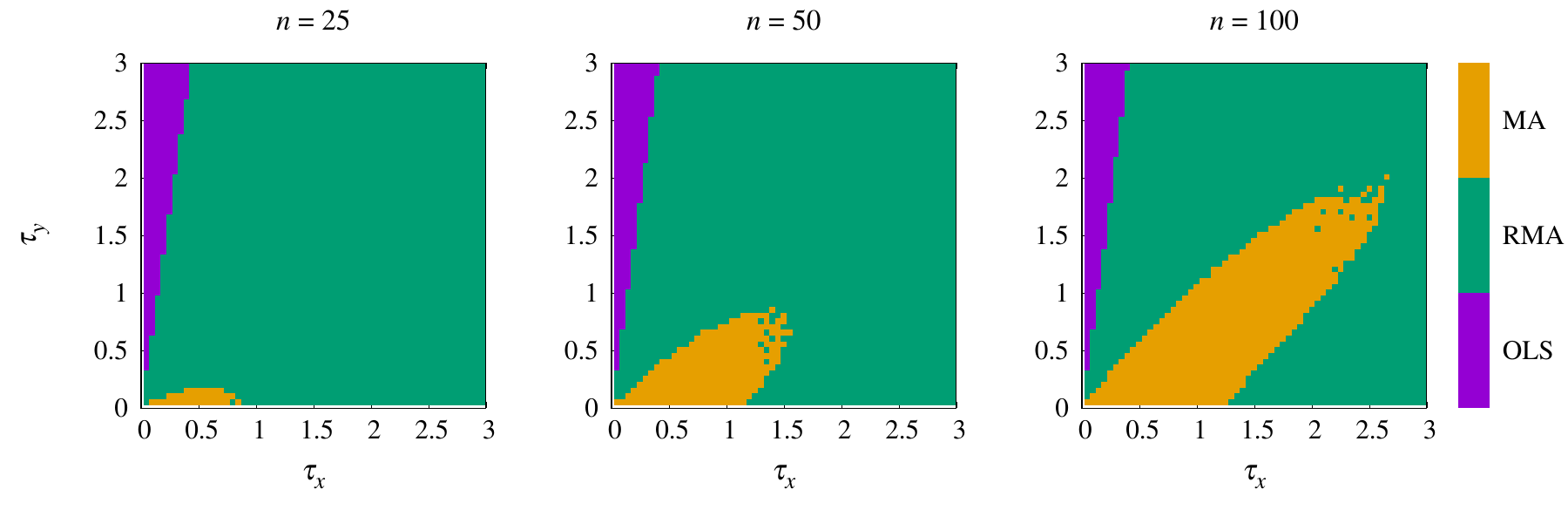}
\caption{Domains of optimality (minimum MSE) for the OLS, MA, and RMA estimators in the $(\tau_x, \tau_y)$ parameter plane for a structural slope $\beta=1.5$. Results illustrate the transition of optimal domains as the sample size increases from $n=25$ (left) to $n=100$ (right). }
\label{fig:4}
\end{figure}   

Figures~\ref{fig:3} and \ref{fig:4} display the phase diagrams for $\beta=0.7$ and $\beta=1.5$, respectively, across sample sizes of $n=25, 50,$ and $100$. These diagrams were constructed using $10^4$ independent Monte Carlo simulations per parameter coordinate to ensure the numerical stability of the dominance boundaries. The results reveal a high sensitivity of the MA estimator to sample size: its region of dominance is nearly non-existent at $n=25$ and expands slowly along the $\tau_x = \tau_y$ diagonal as $n$ increases. However, for large noise variances, the RMA estimator consistently dominates the MA, regardless of whether the noise is symmetric.

Furthermore, the results indicate that the structural slope $\beta$ strongly influences the relative performance of the estimators. For small $\beta$ and low $n$, the OLS estimator dominates a large portion of the phase diagram, particularly in regions of high vertical noise ($\tau_y$). Conversely, for $\beta = 1.5$, the OLS domain is confined to regions of low horizontal noise ($\tau_x$) and remains practically unaffected by the sample size. 

We found that the phase diagrams for $\beta = 1$ (not shown) remain qualitatively similar to their population counterparts. In this specific case, the MA dominance region is totally suppressed even in the population limit—except along the symmetry diagonal $\tau_x = \tau_y$, where it performs identically to the RMA—and this suppression is further exacerbated in finite samples. Because these results do not introduce new qualitative transitions, they are omitted for brevity.

\section{Conclusion}\label{sec:7}

The results presented in this work provide a comprehensive mapping of the trade-offs between the most prominent errors-in-variables estimators. By using the structural framework of Hayes \cite{Hayes_1988} to analytically evaluate biases, we have moved beyond empirical simulation to provide a rigorous, theoretical foundation for estimator selection. Our findings carry significant implications for researchers in both the regression dilution and Regression to the Mean (RTM) communities.

In this work, we focus primarily on the MA and RMA estimators. While these methods were originally designed to preserve geometric symmetry in systems lacking a clear causal hierarchy, our analysis repurposes them as structural tools for handling the specific challenge of measurement error in the independent variable ($X$). Unlike OLS, which assumes the $X$-axis is fixed and error-free, MA and RMA explicitly account for variance in the predictor---the fundamental driver of the RTM effect and slope attenuation in clinical test-retest protocols. By establishing the mathematical properties of these standard geometric estimators, we provide a necessary benchmark to evaluate their effectiveness in correcting artifactual dilution. This approach allows us to determine how they navigate the bias-variance trade-off in finite samples, even in asymmetric scenarios where the researcher’s primary goal is to recover the true biological stability or change between baseline and subsequent measurements.

While OLS, MA, and RMA are sign-agnostic---meaning their mathematical formulation remains valid regardless of whether the structural relationship is positive or negative---other corrections have been developed with a more narrow, directional focus. A prominent example is the estimator proposed by Berry et al. \cite{Berry_1984}, which was designed specifically for clinical test-retest scenarios to account for RTM artifacts in follow-up data. Our derivation of the identity $\beta_B = 1 + \rho(\beta_{RMA} - 1)$ reveals that this method is not a general structural estimator but rather a specialized shrinkage tool. By anchoring the estimate to unity, it intentionally biases the result toward a 1:1 relationship---the clinical null hypothesis that a subject’s state remains unchanged between measurements.

Although this mathematical anchor makes the Berry estimator exceptionally stable and low-variance in clinical settings where $\beta \approx 1$, it renders the method unsuitable for broader scientific application. The most catastrophic failure of this approach occurs when the true structural relationship is negative. In medicine, negative relationships are essential for understanding compensatory mechanisms, such as the inverse scaling between insulin sensitivity and fasting glucose \cite{Kahn_1993}. In ecology, negative slopes are ubiquitous in resource-allocation models, most notably in the Trade-off Hypothesis (TOH) of thermal plasticity \cite{Angilletta_2003,Santos_2025}. The TOH posits a fundamental negative structural relationship between an organism’s thermal breadth and its maximal performance; using a shrinkage estimator anchored to $+1$ in this context would not only dilute the evidence for a trade-off but could mathematically erase the biological signal entirely. Consequently, the Berry method remains a specialized local solution for specific clinical trials, whereas the OLS, MA, and RMA triad provides the global flexibility required for general scientific modeling.

Furthermore, while MA and RMA successfully solve the problem of geometric symmetry by providing a unique regression line, our analysis highlights that in the presence of a clear directional relationship---such as baseline-to-follow-up protocols---the priority shifts from symmetry to reliability. In these asymmetric scenarios, the regression slope is not an ambiguous geometric property but a structural parameter representing biological change or stability. As noted in Equation (\ref{eq:beta_{OLS}}), if the researcher possesses external information regarding the repeatability ($R$) of the independent variable, the true structural slope can be recovered directly from the OLS estimate via the relation $\beta = \beta_{OLS}/R$. As argued in \cite{Fontanari_2026}, providing an educated guess or a sensitivity interval for $R$ based on known technical precision is often more rational than adopting a geometric estimator that assumes equal error distributions. While the MA and RMA solutions are mathematically elegant for symmetric systems, they impose rigid assumptions about the error-variance ratio that may be poorly suited to the directional nature of Regression to the Mean artifacts in clinical research.

Our systematic evaluation of the OLS, MA, and RMA estimators reveals several critical behaviors that dictate their utility in empirical research. First, we find that the performance of OLS is heavily dictated by the magnitude of the structural slope $\beta$. Due to its low sampling variance, OLS remains a powerful tool (in terms of minimum MSE) for small structural slopes and small sample sizes $n$. Crucially, as the OLS estimator is asymptotically unaffected by vertical noise ($\tau_y$), it remains the most reliable choice in scenarios dominated by high measurement error in the dependent variable. However, it is rapidly outclassed by symmetric methods as $\beta$ or $n$ increases. In contrast, the MA estimator is exceptionally sensitive to sample size:  its region of dominance is virtually non-existent at $n=25$ and expands only marginally as $n$ approaches 100. The RMA estimator, however, proves much more robust in finite samples. It dominates large swaths of the $(\tau_x, \tau_y)$ parameter space in high-noise environments, maintaining its superiority  when the noise distribution is asymmetric (specifically when $\tau_x > \tau_y$ for large $\beta$).

Beyond these numerical observations, the primary contribution of this work is the derivation of the population-level analytical slopes. The closed-form expressions  given in Equations (\ref{eq:comp_ols}),   (\ref{eq:comp_rma}), and (\ref{eq:comp_ma}) allow for a rigorous, parameter-free comparison between the estimators. By mapping the boundaries where one method succeeds and another fails, we provide a mathematical framework that replaces heuristic selection with a principled approach based on the expected noise-to-signal ratios of the system under study.

In summary, this work bridges a long-standing gap between the clinical Regression to the Mean literature and the general structural regression framework used in the physical and biological sciences. By situating specialized tools, such as the Berry correction \cite{Berry_1984}, alongside fundamental geometric estimators like OLS, MA, and RMA, we have demonstrated that these seemingly disparate approaches are governed by a single underlying structural identity. Our analytical derivations and MSE phase diagrams provide a unified map for estimator selection, moving beyond the heuristic rules of thumb that often dominate empirical research. For the practitioner, the choice of estimator should not be a matter of disciplinary tradition, but a conscious decision based on the symmetry of the system, the expected direction of the structural relationship, and the prevailing noise environment. 

In navigating these selections, a final note must be made regarding our definition of ``optimality" throughout this work. Because a bivariate system without independent calibration data resides on an inescapable plateau of non-identifiability \cite{Fontanari_2026}, it is impossible to empirically deduce the true ratio of error variances ($\lambda = \nu^2/\delta^2$) from sample moments alone. In a strict sense, the true structural slope can only be recovered flawlessly if $\lambda$ is exactly known. Consequently, when we evaluate the winning or optimal domains of OLS, MA, and RMA in our phase diagrams, we do not present these methods as magical remedies capable of extracting an absolute, unconditioned truth. Instead,   optimality is used here strictly in its decision-theoretic sense: it maps the least vulnerable scheme among three faulty estimators. By illustrating which technique minimizes absolute bias or total Mean Squared Error under varying noise-to-signal coordinates, these diagrams serve as maps of structural sensitivity. They warn the practitioner of the specific mathematical penalties risked when the implicit variance assumptions of a chosen geometric estimator fail to align with the underlying properties of the data, emphasizing that the ultimate benchmark for structural precision remains the empirical quantification of measurement repeatability.

\section*{Acknowledgments}
JFF is partially supported by  Conselho Nacional de Desenvolvimento Cient\'{\i}fico e Tecnol\'ogico  grant number 305620/2021-5.  M.S. is funded by grant PID2024-162000NB-I00 from Ministerio de Ciencia, Innovación y Universidades (Spain).

\appendix

\section{The Deming Estimator: Optimization and Characteristic Equation}\label{appA}
  
\renewcommand{\theequation}{A.\arabic{equation}}
\setcounter{equation}{0}
The Deming regression estimator $\beta_\lambda$ is derived by minimizing the weighted sum of squared residuals in both the $x$ and $y$ coordinates, assuming a known ratio of error variances $\lambda = \nu^2 / \delta^2$ \cite{Deming_1943, Madansky_1959}. Given the observed data points $(x_i, y_i)$, we minimize the objective function
\begin{equation}
S = \sum_{i=1}^n \left[ \frac{(x_i - \hat{X}_i)^2}{\delta^2} + \frac{(y_i - \hat{Y}_i)^2}{\nu^2} \right] = \frac{1}{\nu^2} \sum_{i=1}^n \left[ \lambda(x_i - \hat{X}_i)^2 + (y_i - \hat{Y}_i)^2 \right]
\end{equation}
subject to the constraint $\hat{Y}_i = \alpha + \beta \hat{X}_i$. In this formulation, it is evident that the limit $\lambda \to \infty$ corresponds to the Ordinary Least Squares (OLS) estimator. As $\lambda$ grows, the objective function places an infinite penalty on the horizontal residuals, forcing $\hat{X}_i \to x_i$ and restricting the minimization entirely to the vertical residuals $(y_i - \hat{Y}_i)^2$. Conversely, the choice $\lambda=1$ weights both axes equally, corresponding to the Major Axis (MA) estimator.

By substituting the constraint into $S$ and setting the partial derivatives with respect to the latent variables $\hat{X}_i$ and the parameters $\alpha$ and $\beta$ to zero \cite{Deming_1943, Madansky_1959}, the slope $\beta_\lambda$ is found to be the solution to the following quadratic characteristic equation
\begin{equation} \label{eq:app_characteristic}
\mathrm{cov}(x,y) \beta_\lambda^2 + [\lambda \mathrm{var}(x) - \mathrm{var}(y)] \beta_\lambda - \lambda \mathrm{cov}(x,y) = 0.
\end{equation}
This equation is the origin of the characteristic equation (\ref{car_eq}) used to define the Major Axis (MA) estimator. Specifically, setting $\lambda = 1$ and substituting the population moments $\mathrm{var}(x) = \sigma^2 + \delta^2$, $\mathrm{var}(y) = \beta^2\sigma^2 + \nu^2$, and $\mathrm{cov}(x,y) = \beta\sigma^2$ (normalized by $\sigma^2$ such that $\tau_x = \delta^2/\sigma^2$ and $\tau_y = \nu^2/\sigma^2$) yields the form
\begin{equation}
\beta \beta_{MA}^2 - [(\beta^2 - 1) + (\tau_y - \tau_x)]\beta_{MA} - \beta = 0.
\end{equation}
Solving Eq.~(\ref{eq:app_characteristic}) via the quadratic formula requires an explicit branch selection rule to preserve physical and statistical consistency across both positive and negative structural associations. By systematically selecting the root that preserves the sign of the observed sample covariance $\mathrm{cov}(x,y)$, we obtain the universal, sign-agnostic Deming slope estimator
\begin{equation}
\beta_\lambda = \frac{\mathrm{var}(y) - \lambda \mathrm{var}(x) + \sqrt{[\mathrm{var}(y) - \lambda \mathrm{var}(x)]^2 + 4 \lambda \mathrm{cov}(x,y)^2}}{2 \mathrm{cov}(x,y)},
\end{equation}
which is the equation used throughout this study.


\end{document}